# Granularity of algorithmically constructed publication-level classifications of research publications: Identification of topics

Peter Sjögårde[a,b], Per Ahlgren[c]

[a]Department of ALM, Uppsala University, Uppsala, Sweden
[b]University Library, Karolinska Institutet, Stockholm, Sweden
[c]School of Education and Communication in Engineering Sciences (ECE), KTH Royal Institute of Technology, Sweden

Email: peter.sjogarde@ki.se; perahl@kth.se

Corresponding author: Peter Sjögårde, University Library, Karolinska Institutet, 17177 Stockholm, Sweden

## Abstract

The purpose of this study is to find a theoretically grounded, practically applicable and useful granularity level of an algorithmically constructed publication-level classification of research publications (ACPLC). The level addressed is the level of research topics. The methodology we propose uses synthesis papers and their reference articles to construct a baseline classification. A dataset of about 31 million publications, and their mutual citations relations, is used to obtain several ACPLCs of different granularity. Each ACPLC is compared to the baseline classification and the best performing ACPLC is identified. The results of two case studies show that the topics of the cases are closely associated with different classes of the identified ACPLC, and that these classes tend to treat only one topic. Further, the class size variation is moderate, and only a small proportion of the publications belong to very small classes. For these reasons, we conclude that the proposed methodology is suitable to determine the topic granularity level of an ACPLC and that the ACPLC identified by this methodology is useful for bibliometric analyses.

## 1 Introduction

Classifications of scientific publications have multiple purposes. In libraries, publications can be classified and arranged according to a classification scheme to help users browse a physical collection by subject area.[1] Classifications can also be used within libraries to study circulation statistics or downloads. In the digital world, a classification scheme can be used for information retrieval tasks with the purpose to identify relevant documents for a user, e.g. by refining search results to one or more categories in the classification. Within the bibliometric practice at higher education institutions, classification of research publications can be used to study the structure and processes of research activities and to evaluate research in different subject areas.

Traditional classification schemes used in libraries, such as the Dewey Decimal Classification (DDC) or the Universal Decimal Classification (UDC), were created before the digital era. They were created for shelf arrangement and browsing of physical publications. Each publication was classified manually and placed at the corresponding shelf. The classification was documented on library cards which enabled retrieval of publications by subject area. The granularity of the classification, i.e. how finely or coarsely the classification is grained into classes, had to be set in relation to this physical context. Large, specialized library collections had (and still have) a need for finely grained classifications. Small, general library collections had (and still have) a need for more coarsely grained classifications. The commonly used classification schemes meet these diverse demands by their hierarchical structure. Libraries with large, specialized collections can classify publications at a finely grained level while libraries with small, general collections can use the same classification scheme at more aggregated levels.

---

[1] We use the term "subject area" in a broad sense, to denote an area of research of any level of aggregation. This could be broad areas such as "Computer Science" or more narrow areas such as "Robotic Sensing".

Historically, the physical research journal was classified into classes using the traditional classification schemes. However, individual research publications were not classified, other than assigning them into the same class as the journal issue in which they had been published. This was a natural consequence of the physical media, because publications were physically bound to a journal issue. Today, research publications are born digital and a large proportion of research publications that were published as physical publications the last decades have been digitized. This transition has opened for new possibilities to analyze bibliographic data, which in turn have led to an increased interest in quantitative studies of research publications. As a response to an increased demand for such studies, the research and professional fields of bibliometrics have grown, in particular the last decade. To be meaningful, bibliometric studies commonly require research publications within different broad fields to be classified into narrower areas, and the granularity of the classification is dependent on the purpose of the study.

In our daily practice as bibliometric analysts at a Swedish university, we have regularly received questions from researchers about, e.g. publication quantities, highly cited papers and/or co-publishing. The questions have often been related to specific subject areas, sometimes broad and sometimes narrow, and not uncommonly both; broad to get a comprehensive picture, and narrow to be able to zoom into more finely grained subject areas.

Until a few years ago, the alternatives for subject classification were few. The traditional classification of journals had not been constructed to meet the demands made by the new data analysis practices. These practices require the classification to be comprehensive, uniformly applied through the data collection and to follow a clearly defined set of rules so that the assignment of publications is not dependent on subjective judgements of the classifier.

Alternatives to the traditional classification schemes are applied in the, nowadays web-based, citation indexes. Citation indexes were proposed by Eugene Garfield in 1955, and Web of Science was developed in the 1950s and 60s (Garfield, 1955, 1964). Parallel to the development of the Journal Citation Reports (JCR), where journals are ranked according to citation rates (Garfield, 1972), journal categorization was created (Pudovkin & Garfield, 2002a). The JCR categories were based on similar methods as the classification performed using traditional classification systems, later called a "heuristic procedure" by Pudovkin and Garfield (2002a). More advanced approaches have been proposed for journal classification in recent decades. These approaches use citation relations between journals for their classification (Archambault, Caruso, & Beauchesne, 2011; Boyack, Klavans, & Börner, 2005; Chen, 2008; Doreian, 1988; Leydesdorff, 1987, 2006; Pudovkin & Garfield, 2002b; Rosvall & Bergstrom, 2011; H. G. Small & Koenig, 1977; Zhang, Liu, Janssens, Liang, & Glänzel, 2010; Leydesdorff, Bornmann, & Wagner, 2017).

The many limits of journal-level classification have been acknowledged in the literature (Archambault et al., 2011). An obvious problem is that some journals are broad in scope and thus include publications within different subject areas. Hence, a single subject category cannot accurately represent the subject contents of all publications in such journals. One proposed solution for this problem has been to classify publications appearing in multidisciplinary journals into journal categories created in preceding steps (W Glänzel, 2003; W. Glänzel, Schubert, & Czerwon, 1999; W. Glänzel, Schubert, Schoepflin, & Czerwon, 1999; Gunnarsson, Fröberg, Jacobsson, & Karlsson, 2011). However, this approach solves the problem only partially. In view of this, publication-level classifications are desirable. Considering the high number of publications, manual approaches to publication-level classifications are time consuming and demand enormous amount of resources. Also algorithmically constructed publication-level classifications of research publications (ACPLCs) require a lot of resources, in this case computational resources, much more than journal level classifications. Until recent years, such classifications have been created merely for small or medium size publication sets.

Global[2] subject maps of science have been shown to be more accurate and useful than local maps (Boyack, 2017; Klavans & Boyack, 2011; Rafols, Porter, & Leydesdorff, 2010). Similarly, global classifications have some of the same advantages. For example, they may be useful for studies (a) where subject differentiation is of importance, (b) dealing with identification and analysis of emerging research fields (Milanez, Noyons, & de Faria, 2016; H. Small, Boyack, & Klavans, 2014), and (c) aiming to reveal relations between subject areas. Local, small or medium scale mappings or classifications do not provide the same possibilities. To facilitate such studies, global publication-level classifications have been constructed in recent years (Boyack & Klavans, 2014a, 2014b; Šubelj, van Eck, & Waltman, 2016; Waltman & van Eck, 2012, 2013a). This development is a huge step forward in the area of research classification. Nevertheless, the methods for ACPLCs are in need for development. In this article, we will address one of the challenges that hitherto have been addressed only briefly.

The issue that we deal with in this paper is how to set the resolution parameter for cluster solutions at the level of research topics in an ACPLC that involves a parameter of this kind, i.e. to determine, in such an ACPLC, the granularity of the classification at this hierarchical level. So far, this has not been a topic much discussed in the literature. Waltman and van Eck mention that "the choice of parameter values should be guided by the purpose for which a classification system is intended to be used." (2012, p. 2383) Boyack and Klavans have focused on which citation relation to be used (Boyack et al., 2011; Klavans & Boyack, 2017), rather than the granularity of the classification. Similarly to Waltman and van Eck, Boyack and Klavans point out that the "proper level of granularity likely depends on the specific question being asked, and is a question that we do not address in this study." (Klavans & Boyack, 2017, p. 994)

The purpose of this paper is to find a theoretically grounded, practically applicable and useful granularity level of an ACPLC with respect to topics. We plan to address the level of research specialties in future research.[3] To determine the granularity of topics, a baseline classification is constructed. Synthesis papers and their references are used to create a baseline classification. ACPLCs with different granularities, constructed by the use of different values of the resolution parameter, are then compared to the baseline classification. The classification, with its corresponding resolution parameter value, that best fit the baseline classification, is proposed to be used for the bibliometric analysis of topics.

This remainder of this paper is structured as follows. The next section (2) contains the framework of the study. We outline the state of research related to the construction of ACPLCs and discuss the topic notion. In Section 3, data and methods are described. The results are reported and discussed in Section 4. In the last section (5), conclusions are put forward.

## 2 Framework of the study

We agree with others (Wolfgang Glänzel & Schubert, 2003; Gläser, Glänzel, & Scharnhorst, 2017; Klavans & Boyack, 2017; Mai, 2011; Smiraglia & van den Heuvel, 2013; Velden et al., 2017; Waltman & van Eck, 2012) that there is no one perfect classification that can be used for all purposes and that the methods used to obtain a classification of research publications should be guided by the purpose of the use of the classification. Nevertheless, we believe that a wide range of bibliometric studies on topics (as well as on other levels of hierarchy, e.g. research specialties) have similar purposes. Therefore, we think that there is a need to create a best practice for obtaining ACPLCs. Further, there is a need for a common understanding of what we refer to by the term "topic" and how topics can be related to, and identified by, the classification of publication collections. How large or small is a topic? So far, the size of topics may vary between different studies, making comparisons of studies problematic or inappropriate. A common definition and a standardized approach for finding granularities of publication classes corresponding to topics improves the possibilities for comparison between studies. Further, when working in a bibliometric practice, there is a need to create classifications that can be updated and used recurrently for efficient bibliometric analyses. A standard classification of publications corresponding to topics would be of great use in such context.

---

[2] "Global" refers to a comprehensive coverage of subject areas. Similarly, "local" refers to the coverage of one or a few related subject areas.

[3] We use the American English version of the term "specialty". The British English version "speciality" is sometimes used in the literature.

## 2.1. Algorithmic classification

Algorithmic classifications can be created by community (cluster) detection techniques, like modularity- or flow-based techniques (Fortunato, 2010), or by more traditional techniques, like *k*-means clustering. Community detection techniques cluster vertices (or nodes) related to each other by edges (or links) in a network. Partitions, i.e. cluster solutions, are created so that vertices within a cluster are more strongly related to each other than to vertices outside the cluster. The term "clusters" are sometimes used in the literature to denote the members of the resulting partitions. However, since our goal is to create a classification, we find it natural to use the term "classes", and this term is used in the remainder of this article. The ACPLCs that we work with in this study constitute output from the program Modularity Optimizer[4], created by Ludo Waltman and Nees Jan van Eck (2013a), and in which two modularity functions are implemented (Newman & Girvan, 2004; Traag, Van Dooren, & Nesterov, 2011), together with algorithms for optimization of the functions, like the smart local moving (SLM) algorithm (Waltman & van Eck, 2013a). The implementation made in this study is based on the methodology put forward in Waltman & van Eck (2012). The SLM algorithm makes use of a resolution parameter, and is thereby able to detect communities at different levels of granularity. The software includes two different modularity functions. We used the alternative function (Traag et al., 2011). Further, the methodology includes a relatedness measure with respect to pairs of publications, a measure that normalizes for differences in citation volumes between fields, caused by the different reference practices (Waltman & van Eck, 2012). This kind of normalization is essential, because fields with a high number of references per paper otherwise would have greater density in the network, and fields with fewer references, and therefore less density, could be incorporated in these higher density fields.

As noted by Šubelj et al. (2016), an approach including a resolution feature "requires a careful choice of parameter values." With this paper, we attempt to contribute to this choice of the resolution parameter at the granularity level of topics.

In bibliometric publication-level networks, vertices represent publications, whereas edges usually represent direct citations, bibliographic coupling (Kessler, 1965), co-citations (e.g. Marshakova-Shaikevich, 1973; H. Small, 1973), textual similarity (e.g. Ahlgren & Colliander, 2009; Boyack et al., 2011) or combined approaches (e.g. Colliander, 2015; Wolfgang Glänzel & Thijs, 2017).

Referencing is a communicative (at least partly) practice (see Moed, 2005 chapter 15, for a discussion on what references and citations measure) and citation-based networks between publications express formal communication taking place within the research community. References also represent an expression of the cognitive structure of the community; a researcher citing a paper is obviously familiar with the cited paper and relates to that paper in her or his research. Co-citations has been used to capture the intellectual structure of a research community and bibliographic coupling for representing the research front.

Textual similarity between documents expresses relations of a rather different nature, notwithstanding such relations are likely to co-occur with citation relations. It expresses topic similarity in a more direct way, which may be a strength depending on the purpose of the classification. However, the textual similarity approach has some disadvantages. The approach is more complex, more computationally demanding and may add noise into the similarity measure.

---

[4] http://www.ludowaltman.nl/slm/

The traditional classification of publications, based on perceived subject similarity by the classifier, lacks relations between publications within a class and does neither express the communication taking place nor the cognitive structure. Hence, algorithmic classification does not only enable efficient large scale classification of publications, moreover, it provides opportunities to analyze the communicational and cognitive structure of research.

A thorough discussion of different publication-publication similarity measures is out of the scope of this paper. However, we acknowledge that such discussion is of great importance for the development of a standard methodology for creation of ACPLCs. However, our approach to set the granularity levels of ACPLCs is not delimited to a particular publication-publication similarity measure, but can be used in combination with any such measure.

Direct citations are used in this study for two reasons. (1) Direct citations give rise to fewer relations than e.g. co-citations or bibliographic coupling, making it possible to create an ACPLC on larger datasets. This is important, since global, both in terms of subject representation and in terms of geographical uptake, citation databases are large in publication volume. For example, the order of magnitude of the amount of articles from 1980 to 2016 in Web of Science is currently around 30 million. Between those articles there are about 600 million direct citation relations. The number of bibliographic coupling and co-citation relations are much larger. This number can be approximated to about 100 billion for bibliographic coupling and around 30 billion for co-citations.[5] Efficiency of the algorithms for constructing ACPLCs is therefore of great importance. (2). There is empirical support that direct citations performs well in comparison with bibliographic coupling as well as co-citations when it comes to larger datasets.[6] A recent study used concentration of references from articles with at least 100 references and textual coherence to evaluate the outcome of cluster solutions based on bibliographic coupling, co-citations and direct citations (Klavans & Boyack, 2017). The authors discovered that, if larger time frames are used, direct citations perform better than bibliographic coupling and co-citations. For this reason, the authors propose that direct citations should be used for the creation of taxonomies of science. Even if we use direct citations in this study, we believe that the choice of publication-publication similarity measure needs further research.

---

[5] Let $N$ the number of source publications in the database, and $C_i$ ($R_i$) the number of citations (cited references pointing to source publications) to (of) the *i*th source publication. Then the number of *bibliographic coupling relations* in the database is equal to

$$\sum_{i=1}^{N} \frac{C_i(C_i - 1)}{2} \qquad (1)$$

whereas the number of *co-citation relations* in the database is equal to

$$\sum_{i=1}^{N} \frac{R_i(R_i - 1)}{2} \qquad (2)$$

(If bibliographic coupling takes non-source publications into account, $N$ in Eq. (1) stands for the number of unique cited references in the database, and $C_i$ for the number of source publications that cite the *i*th cited reference. If the co-citation analysis takes non-source publications into account, $R_i$ in Eq. (2) stands for the number of cited references of the *i*th source publication.)

[6] However, bibliographic coupling might be preferable for small or medium size datasets (Boyack & Klavans, 2010; Waltman, Boyack, Colavizza, & van Eck, 2017).

## 2.2. Research topics

Methods to detect and map topics have been developed within the disciplines of information retrieval, bibliometrics and computational linguistics (see Velden et al., 2017 for a comparison of topic extraction approaches). This has been done by the use of citation relations between documents (Boyack, Klavans, Small, & Ungar, 2014; H. Small et al., 2014; Upham & Small, 2010) or term relations within or between documents or sets of documents (Besselaar & Heimeriks, 2006; M. Callon, Courtial, & Laville, 1991; Michel Callon, Courtial, Turner, & Bauin, 1983; Leydesdorff & Nerghes, 2016; Song, Heo, & Kim, 2014; Wang, Cheng, & Lu, 2014; Wang et al., 2014; Yan, 2014; Yan, Ding, & Jacob, 2012; Yan, Ding, Milojević, & Sugimoto, 2012). However, the term "research topic" is not well defined in the research literature and the term is often used without definition, sometimes synonymously with other terms such as "research area" or "subject area".

Two issues that have not been studied thoroughly in the literature are the granularity of topics and the operationalization of the notion. For instance, in an interesting study, Milanez et al. (2016) study topics within nanocelluloses using the ACPLC developed by Waltman and van Eck (2012, 2013a). However, the terms "research topic" and "research area" are used synonymously and without definition. The authors further consider classes at the lowest hierarchical level as representing topics, and thereby (implicitly) assume that the value of the resolution parameter for this level gives rise to topics. Small et al. (2014) study emerging topics by the use of an ACPLC. The resolution parameter used to obtain the ACPLC is set arbitrarily, so far as we can see. The authors briefly acknowledge that some of the topics identified in the study might be considered as sub-topics. However, they do not discuss the problem of setting the resolution parameter further. A third example can be found in Yan et al. (2012). This work explores the relation between topics and communities, using two different clustering techniques: one modularity-based and *k*-means clustering. Referring to Blei and Lafferty (2007), the authors give the following definition: "[a] topic represents an underlying semantic theme and can be informally defined as an organization of words and can be formally defined as a probability distribution over terms in a vocabulary" (Yan, Ding, & Jacob, 2012, p. 500) . However, the granularity of the corresponding operationalization is not discussed and the value of the resolution parameter is not reported.

Another approach to identify topics is to use search terms to retrieve a publication set, which is considered to constitute a topic (Kiss, Broom, Craze, & Rafols, 2009). Such techniques may be useful in some cases. However, besides being time consuming, search terms may be broader or narrower than the scope of the topic, and some topics will be harder to define by search terms than others. If one would like to, for example, compare the growth and spread of one topic with other topics, this may be difficult because of the method's inherent inconsistency with regard to publication retrieval for the different topics and differentiation between topics. The search terms used for the identification of publications within one topic may be broader or narrower than the search terms used to identify another topic. This problem is similar to the resolution problem when using ACPLCs.

### 2.2.1. Explication of the research topic notion

We agree with Yan et al. (2012) that a topic "represents an underlying semantic theme". Further, a topic corresponds to a problem area addressed by researchers. As such it includes a set of research questions addressed by one or several research communities. In agreement with van den Besselaar and Heimeriks (2006), we see topics as the lowest level of aggregation to be considered for classification of subject areas. Examples of topics within the field of scholarly communication research is: (1) h-index and similar researcher level indicators, (2) journal indicators, (3) open source and open access in scholarly communication, and (4) the peer review process.

We acknowledge that the delineation of topics is complex. Topics can be overlapping, addressed by several research specialties (Yan, Ding, & Jacob, 2012; Yan, Ding, Milojević, et al., 2012), shift in focus and vocabulary over time and vary in size (in terms of publications and number of researchers addressing the topic).

In concordance with Klavans and Boyack (2017), we consider topic as the subject area level that is, in general, addressed by researchers in review publications. Review publications typically summarizes the background and current state of the research conducted within a problem area. Thereby each such paper can be seen as a synthesis of a topic. As noted by Klavans and Boyack, there is no common definition and operationalization of review publications. For this reason, they define synthesis papers "as those with large numbers of references, regardless of their database designation as an article or review." In this paper, we use the same definition of synthesis papers. Table 1 lists four examples of synthesis papers, and their corresponding topics, within the field of scholarly communication research.

| Topic | Synthesis paper |
|---|---|
| h-index and similar measures | Zhang, L, Thijs, B, Glänzel, W. (2011). The diffusion of H-related literature. Journal of Informetrics, 5(4), 583-593. |
| Journal indicators | Vanclay, JK. (2012). Impact factor: outdated artefact or stepping-stone to journal certification?. Scientometrics, 92(2), 211-238. |
| Open source/access | Aksulu, A, Wade, M. (2010). A Comprehensive Review and Synthesis of Open Source Research. Journal of The Association for Information Systems, 11(11), 576-656. |
| Peer review | Souder, L. (2011). The ethics of scholarly peer review: a review of the literature. Learned Publishing, 24(1), 55-72. |

Table 1: Example of topics and corresponding synthesis papers within the field of scholarly communication research.

## 2.3. Model of an ACPLC

In this section we present a network model of an ACPLC. Figure 1 visualizes an instance of the model and shows publications (nodes), relations (edges) and how the publications are classified into classes at two hierarchical levels (represented by colors). Edges represent any publication-publication relation such as direct citations, bibliographic coupling, co-citations or textual similarity. We delimit this model to two levels of hierarchy, topics and specialties. However, more levels can be added to the model. Furthermore, the visualized instance exemplifies how publications published by a researcher (the corresponding nodes have red borders) can belong to different topics and different specialties. The model comprises a logical classification: Each publication is classified into exactly one class at each level of hierarchy.[7] Moreover, all publications in a class, at a level below the top level, are classified into exactly one, and the same, parent class. It follows that each topic in the model belongs to exactly one specialty. This is a shortcoming of the model, since theoretically, topics can be addressed by several specialties (Yan, Ding, & Jacob, 2012) or, at a higher level of aggregation, disciplines (Wen, Horlings, van der Zouwen, & van den Besselaar, 2017). However, a global classification of research publications that have the purpose to be used recurrently for e.g. compilation of statistics, regarding publication output within different subject areas, have some practical requirements. Compiled statistics of this kind need to be easily interpretable by others than bibliometric specialists. Overlapping classes are harder to interpret and often require fractionalization when statistics are to be compiled. Further, and importantly, the relation between a topic and other specialties than the parent specialty, as well as relations between topics, can still be expressed and analyzed by use of the relational strengths associated with the edges in the model.

---

[7] A *logical classification* of a set of objects, $O$, is a set $C$ of non-empty subsets of $O$ such that (a) the union of the sets in $C$ is equal to $O$, and (b) the sets in $C$ are pairwise disjoint. Thus, each object in $O$ is classified into exactly one set in $C$.

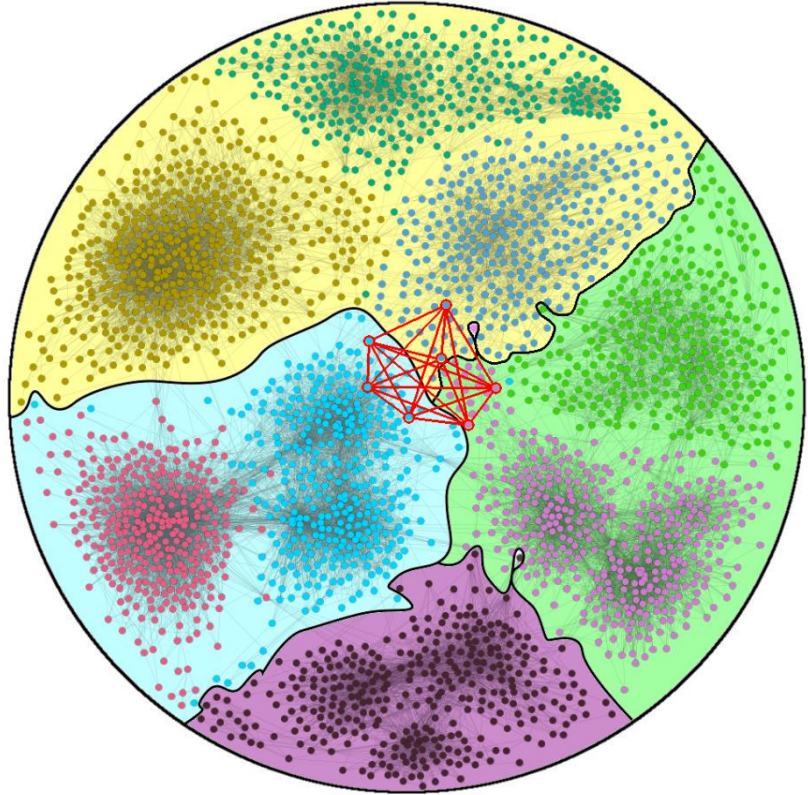

*Figure 1: Instance of a model of an ACPLC at the level of topics and specialties. Nodes represent publications, whereas edges represent any publication-publication relation such as direct citations, bibliographic coupling, co-citations or textual similarity. Nodes with red borders represent publications authored by a given researcher. Nodes are colored according to their topic belonging. Background color indicates the belonging of publications (and topics) at the level of specialties.*

A best practice regarding which publication-publication similarity measure to be used for large-scale classification of research publications, and to be used for a standardized procedure to create ACPLCs, would be useful for bibliometric practices. We do not consider the issue of which publication-publication similarity measure to be used as fully answered by the literature, and therefore we identify it as important for future work. In the model, relations between nodes can be expressed by any of the above-mentioned publication-publication relations. Another issue that we do not address in this paper is whether or not to include non-source publications, i.e. "references for which an indexed source record does not exist in the database" (Boyack & Klavans, 2014b). Such an approach may add robustness to the methodology and may therefore be preferred in a standardized procedure for the creation of ACPLCs.

## 3 Data and methods

KTH Royal Institute of Technology's bibliometric database Bibmet was used for the study. Bibmet contains Web of Science publications from the publication year 1980 onwards. Publications registered in Web of Science at the time for data extraction (March 2017), and of the Web of Science document types "Article" and "Review", were included in the study, a total of 33,073,303 publications from 1980-2017.[8] Of these, 2,403,938 had no citation relation to any other publication in the publication set. These articles were excluded from the study.[9] Thus, 30,669,365 publications remained and constitute the publication set of the study. Let *P* be this set. In the remainder of this paper, we use the term "article" to refer to both articles and reviews.

## 3.1. Design of the study

We attempt to find a granularity of an ACPLC, where the ACPLC is based on the articles in *P*, that correspond to topics. In order to identify the granularity of topics, a baseline classification of publications (BCP) is created. The BCP is a set of publications, where the publications are considered as classes, and each member of a class in BCP is a publication referred to in the reference list of the class, i.e. of the publication. The BCP is compared to several ACPLCs with different granularities. An appropriate granularity is detected and an ACPLC is chosen, the classes of which correspond to topics. The methodology is described in detail in step I to IV below and schematically illustrated in Figure 2.

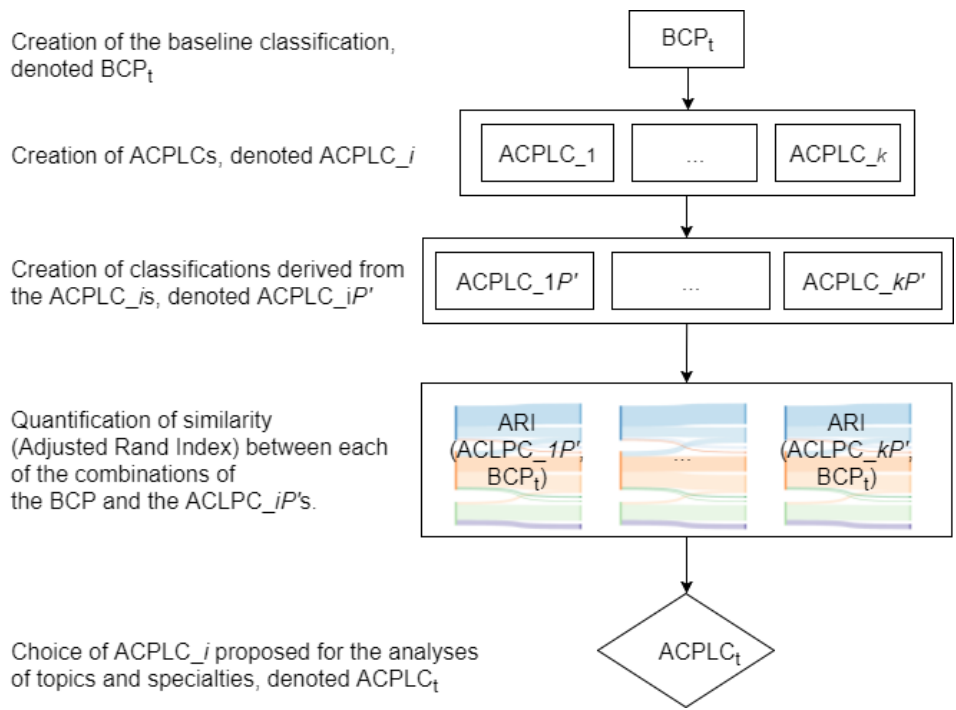

*Figure 2: Illustration of the design of the study.*

### I. Creation of baseline classes

We construct a baseline classification to correspond to topics, which we denote by $BCP_t$. For the creation of $BCP_t$, we use synthesis articles in *P*, operationalized as articles with at least 100 references in correspondence to the approach developed by Klavans and Boyack (2017). Each such article constitute a class, and its list of cited references points to the reference publications of the class. The reason for operationalizing synthesis papers as articles with 100 references or more is well-motivated by Klavans and Boyack. Such articles are to a high degree classified as reviews and authored by influential authors. They are more highly cited than articles with less than 100 references. Further, their reference publications are widely distributed across subject areas and the overlap of reference publications (publications cited by more than one of the synthesis papers) is small, indicating that they do not treat the same subject area.

---
[8] The publication years 2016-2017 were not completely registered at this point in time.
[9] 1.2 million of the excluded publications have no references. A higher share of the excluded publications have publication years in the beginning of the time period than in the end of the period. E.g., there are about 123 thousand publications excluded from the publication year 1980 and 37 thousand publications excluded from the publication year 2015.

Because $BCP_t$ is to be used as a baseline to estimate granularity of an ACPLC with respect to topics, there are some requirements on its properties:

A. To be able to compare the classifications, the union of the classes in $BCP_t$ must be a subset of the union of the classes in an ACPLC.
B. Ideally, each class in $BCP_t$ should address exactly one topic and each pair of distinct classes should address different topics.
C. The classes in $BCP_t$ must not be overlapping. Hence, a reference publication should only belong to one class.

In order to satisfy point A, we restricted the reference publications to articles belonging to *P*. Thus, all cited references that were used to construct $BCP_t$ are *active* references: references that point to publications covered by the data source (Waltman et al., (2013)).

For the former part of point B (each class in $BCP_t$ should address exactly one topic), this has to some extent been dealt with by the use of synthesis articles as classes in our baseline classification, since such an article roughly treats a certain topic. Regarding the latter part of point B (each pair of distinct classes should address different topics) and point C, we proceeded as follows. To lower the risk of obtaining multiple synthesis articles covering the same topic, we delimited them to articles in *P* published in one year only, namely year 2015. Since Bibmet covers articles from 1980 onwards, the included articles have a 35 year window for their reference articles to be included in the study. Further, we only included synthesis articles with at least 80% active references (Klavans and Boyack, (2017)). After these limitations, 37,476 synthesis articles remained. Still, this set of articles was likely to contain articles addressing the same topic. Bibliographic coupling was used to determine if the remaining articles did contain articles addressing the same topic. If two articles had an overlap of 30% or more regarding their active reference articles, they were considered as topic overlapping.[10] The level was set after some testing and subsequent examination of the results. The threshold, which is quite low, was set in order to avoid to obtain more than one synthesis article addressing the same topic. We grouped articles so that all articles that were directly or indirectly connected, by an active reference article overlap of 30% or more, were assigned the same group. E.g. if synthesis article $s_1$ has an active reference article overlap of ≥ 30% with article $s_2$, and $s_2$ has an active reference article overlap of ≥ 30% with $s_3$, then $s_1$, $s_2$ and $s_3$ are assigned to the same group. Note that $s_1$ and $s_3$ are assigned to the same group, even if they do not have an active reference article overlap of ≥ 30%. Each obtained group of articles was considered as addressing the same topic. One of the articles was then randomly selected from each group.[11] This procedure excluded 2,012 of the 37,476 (about 5%) synthesis articles.

---

[10] The overlap (*y*) between two synthesis papers ($s_1$ and $s_2$) is given by:

$$y = \frac{1}{2}\left(\frac{m}{A_1} + \frac{m}{A_2}\right) \quad (3)$$

where *m* is the number of shared active reference articles, i.e. active reference articles occurring in both $s_1$ and $s_2$, $A_1$ the number of active reference articles in $s_1$ and $A_2$ the number of active reference articles in $s_2$. Note that we give the overlap measure threshold as a percentage in the running text.

[11] An alternative approach would be to merge the reference articles in the group and consider this list of reference articles as one baseline class. We tested this approach, and the result was a slight increase of the ARI values. However,

If the synthesis articles still had overlapping active reference articles after the procedure, i.e. articles in *P* cited by more than one of the remaining synthesis articles, the overlapping reference articles were assigned to exactly one of their synthesis articles (and erased from the other ones). This assignment was based on the bibliographic coupling strength (i.e., the number of shared references) between the overlapping reference article and the other active reference articles of its synthesis articles. Let $s_1, ..., s_m$ ($m \geq 2$) be the synthesis articles of an overlapping reference article *a*. For each $s_i$ ($1 \leq i \leq m$), the bibliographic coupling strength between *a* and each other, relative to *a*, active reference article of $s_i$ was calculated. Then the sum of the coupling strengths across these other active reference articles of $s_i$ was calculated, which yielded a similarity value with respect to *a* and $s_i$. Finally, *a* was only kept in the reference list of that $s_i$, with which *a* had the highest similarity value. In case of ties, *a* was erased from all *m* reference lists. After the assignments in question, point C was satisfied, whereas the latter part of point B can be assumed to have been satisfied to a large extent.

In total, $BCP_t$ contain 35,464 synthesis articles (classes) and 2,786,203 reference articles. We denote the union of the classes in $BCP_t$ as *P'*.

## II. Creation of ACPLCs of different granularity with respect to the topic level

To obtain ACPLCs of different granularity, we used the program Modularity optimizer, setting the resolution parameter to different values. Normalized direct citation values between the publications in *P*, as proposed by Waltman and van Eck (2012), were given as input to Modularity optimizer, a total of approximately 614 million edges. By this, ACPLCs were created for comparison of similarity with $BCP_t$. We denote the ACPLCs by ACPLC_1, …, ACPLC_*k*, where *k* is the number of created ACPLCs.

## III. Creation of classifications derived from the ACPLCs

For each ACPLC_*i* ($1 \leq i \leq k$), a classification was derived from ACPLC_*i* in the following way:

(a) Each class *C* in ACPLC_*i* such that *C* did not contain any articles in *P'* was removed from ACPLC_*i*. Let ACPLC_*i*1 be the subset of ACPLC_*i* that resulted from the removal.
(b) For each class *C* in ACPLC_*i*1, all articles in *C* that did not belong to *P'* were removed from *C*. Let ACPLC_*iP'* be the set that resulted from these removal operations.

Clearly, the set ACPLC_*iP'* constitutes a (logical) classification of *P'*, i.e. of the union of the classes of the baseline classification $BCP_t$. Thus, ACPLC_*iP'* and $BCP_t$ have exactly the same underlying reference articles. Notice, however, that ACPLC_*iP'* is not a subset of ACPLC_*i*, the algorithmically constructed classification of articles in *P* from which it was derived.[12] We denote the *k* derived classifications as ACPLC_1*P'*, …, ACPLC_*kP'*. These classifications then correspond to the classifications ACPLC_1, …, ACPLC_*k*.

## IV. Quantification of the similarity between $BCP_t$ and the ACPLC_*iP'*s

We attempt to optimize the granularity of an ACPLC_*iP'* so that it exhibits as high similarity as possible with $BCP_t$. Figure 3 illustrates the relation between two classifications as an alluvial diagram. Example *A* shows two classifications $A_1$ and $A_2$ with a high similarity. Example *B* shows two classifications where one of the classifications is more coarsely grained ($B_1$) than the other classification ($B_2$). The similarity between $A_1$ and $A_2$ is higher than the similarity between $B_1$ and $B_2$. If we consider $B_1$ as a baseline classification, then the granularity of $B_2$ would be too finely grained.

---

since the merged approach would violate the definition of a baseline class, we chose to use a randomly selected synthesis article.

[12] Even if this is theoretically possible, though: ACPLC_*iP'* is a subset of ACPLC_*i* if and only if for each class *C* in ACPLC_*iP'*, the class in ACPLC_*i*1 that *C* is obtained from is identical to *C*. If the latter is the case, all articles in *P* not belonging to $BCP_t$ belong to the classes in ACPLC_*i* that are removed in step (a) above.

To quantify the similarity between BCP$_t$ and an ACPLC_$iP'$, the Adjusted Rand Index (ARI) (Hubert & Arabie, 1985) was used. The ARI ranges from 0 to 1. It is advantageous over the original Rand Index proposed by Rand (1971), because it adjusts for chance. The ARI compares two classifications by considering pairs of items in one of the classifications and whether or not each pair is grouped into the same class in the other classification. Item pairs that are grouped into the same class in both of the compared classifications increase the ARI value. Pairs that are not grouped into the same class in neither of the two classifications also increase the value. In the appendix, ARI is defined.

In contrast to the Herfindahl index, as implemented by Klavans and Boyack (2017), the ARI decreases the value if a pair of objects in a classification are grouped together within one classification but are separated in the other. This feature of the ARI is essential for our study. The Herfindahl index approach, as implemented by Klavans and Boyack, gives an optimal value if all publications in an ACPLC are assigned to one single class. In such a case, every pair of reference articles in *P'* would belong to the same class in the ACPLC. Obviously, this would not be a granularity of the ACPLC that corresponds to topics. Note that an ARI value of 1 between BCP$_t$ and an ACPLC_$iP'$ corresponds to a situation in which these two classifications are identical.

To find the ACPLC_$iP'$ with the highest ARI similarity with BCP$_t$, we tested the similarity after each run of Modularity optimizer. A first run was made with a resolution parameter value of 0.00005. This value was chosen based on previous experience and some testing. We then increased the parameter value with 0.00005. This increase resulted in a higher ARI similarity, and we therefore increased the resolution further with 0.00005 for the third run, from 0.00010 to 0.00015. We continued by increasing the resolution by 0.00005 until the ARI value decreased, in total three more times, and thus six runs were done. The third run, with a resolution parameter value of 0.00015, gave rise to the highest ARI similarity (Table 2 and Figure 4 in Section 4).

Since a given ACLPC_$iP'$ consists of 2,786,203 articles, covering almost 9% of the articles in the corresponding ACPLC_$i$, we anticipate this selection to be representative of ACPLC_$i$. The ACPLC_$i$ such that ACLPC_$iP'$ exhibits the largest ARI similarity with BCP$_t$ is proposed to be used for the analyses of topics. We denote this ACPLC_$i$ by ACPLC$_t$.

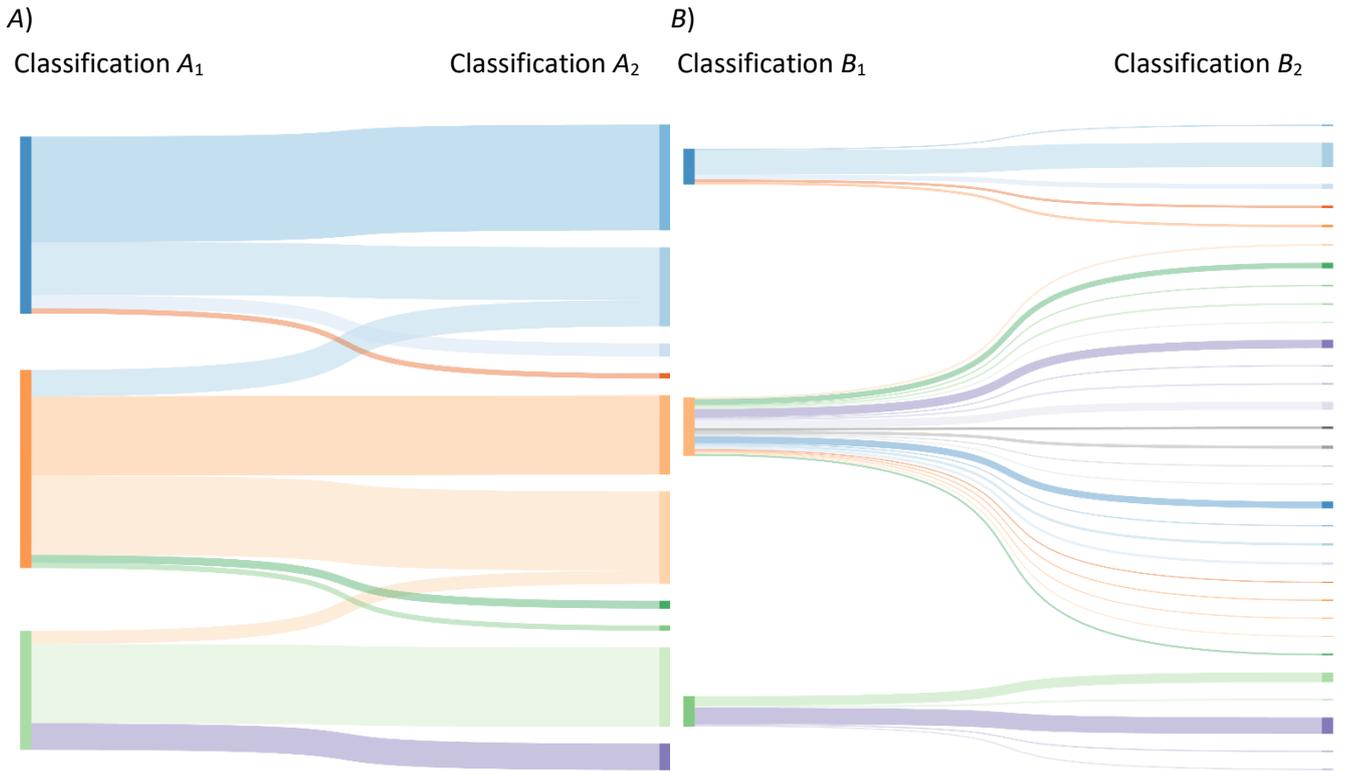

*Figure 3: Two alluvial diagrams (A and B) illustrating the relation between two classifications. A shows two classifications with a high level of similarity. B shows two classifications with a low level of similarity.*

## 4 Results and discussion

In this section, we first deal with the selection and the properties of ACPLC_$i$ corresponding to ACPLC_$iP'$ with the highest ARI similarity to BCP$_t$ (denoted ACPLC$_t$). We discuss the class size distribution of ACPLC$_t$ and the validity of the results. Thereafter, we use two cases to explore if the topics obtained by the methodology used in this study make intuitive sense (Šubelj et al., 2016). More precisely, we study the topics of (1) the articles published in *Journal of Informetrics* (JOI) and (2) the active reference articles of a review article within nanocelluloses.

### 4.1. Selection and properties of ACPLC$_t$

Figure 4 shows a scatter plot of the relation between the resolution value (horizontal axis) used to obtain ACPLC_$i$s and the ARI value (vertical axis), obtained by comparing the ACPLC_$iP'$'s with BCP$_t$. The data points in Figure 4 form a slightly skewed negative parabola shaped curve. The data point at the top of this curve, having the highest ARI value, corresponds to ACPLC_3$P'$, which in turn corresponds to ACPLC_3. Consequently, we consider ACPLC_3 to be the most proper ACPLC_$i$ with respect to granularity and topics. In the remainder of this paper, we denote ACPLC_3 as ACPLC$_t$. However, the slopes of the top part of the curve are gentle. The ARI value changes only slightly if the value of the resolution parameter shifts from 0.00015 by e.g. 0.00005 in either direction. Thus, ACPLC_2$P'$ and ACPLC_4$P'$ perform almost as good as ACPLC_3$P'$.

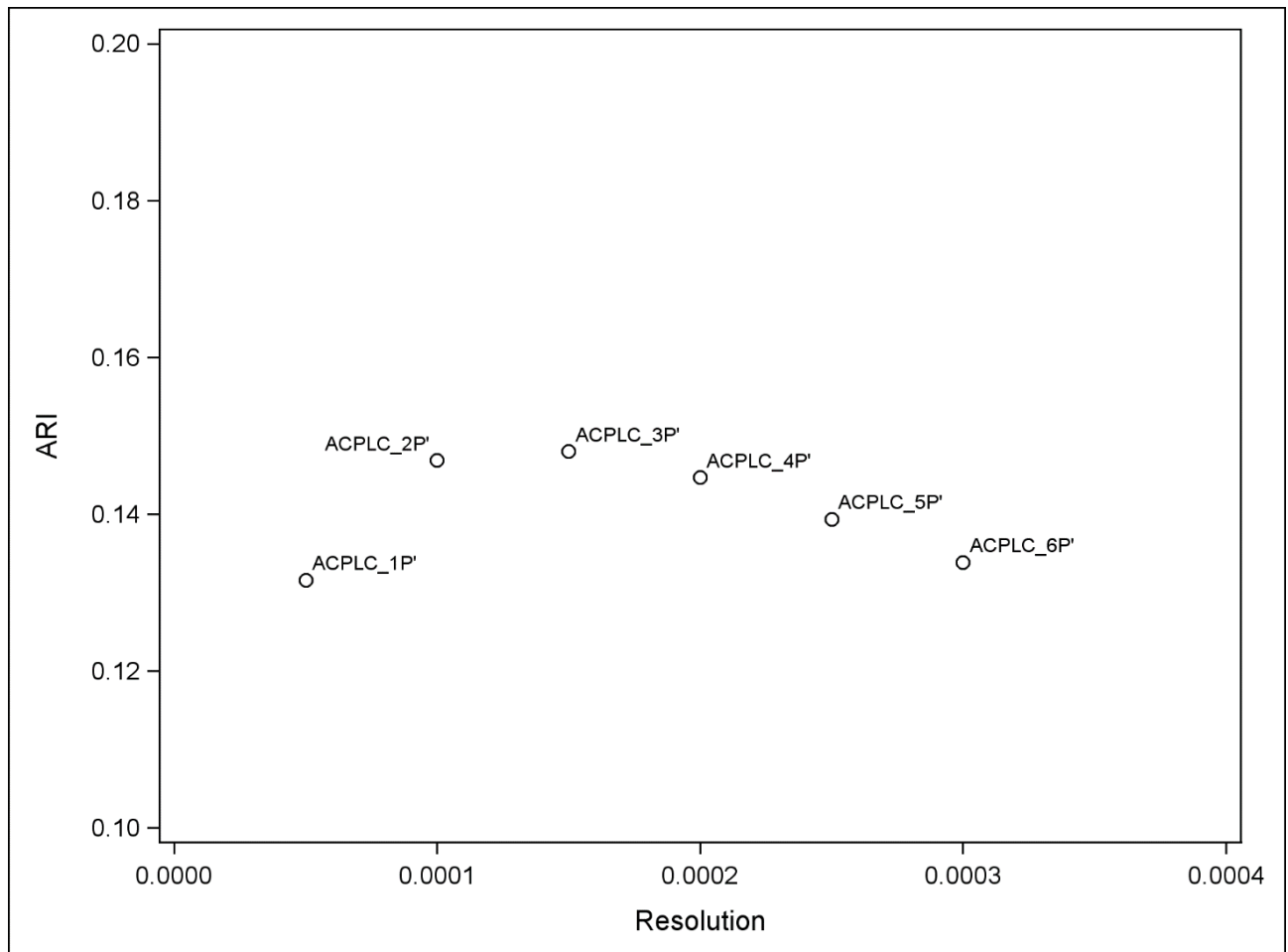

*Figure 4: ARI values between ACPLC_iP's and $BCP_t$. The vertical axis shows the ARI value and the horizontal axis shows the value of the resolution parameter used to obtain the corresponding ACPLC_is. The order of ACPLC_iP's corresponds to their order in Table 2.*

How well does $ACPLC_t$ match $BCP_t$? This question is not easy to answer. The ARI value is one aspect and does not say much about how reference articles in classes in $BCP_t$ are distributed into classes in $ACPLC_t$. Further, ARI values vary depending on the type of data that is being analyzed. This property makes it hard to estimate if an ARI value should be considered as high or low.

An option to illustrate the similarity between $BCP_t$ and $ACPLC_t$ is to use an alluvial diagram, as exemplified in Figure 3. However, since the data set is large, it is impossible to get a comprehensive picture from an illustration of the whole data set. For this reason, we have created an alluvial diagram based on the distribution of the reference articles of an average $BCP_t$ class into classes in $ACPLC_t$. This was done by first calculating the average number of classes in $ACPLC_t$ into which the reference articles in a class in $BCP_t$ are distributed, an average that is equal to 29 (after rounding to nearest integer). We then selected all 801 classes in $BCP_t$ that were distributed into exactly 29 classes. Let the set of these classes be $P_{tc}$. The average number of reference articles in a $P_{tc}$ class is 73.5. For each of the $P_{tc}$ classes, we calculated the number of its articles in each of the 29 $ACPLC_t$ classes and sorted the resulting table in descending order. The $ACPLC_t$ class with the highest number of articles (i.e. the class corresponding to the first row in the table) was assigned the rank 1, the second largest class (i.e. the class corresponding to the second row in the table) was assigned the rank 2, etc. In this way, 801 ranked tables were obtained. Finally, averages of the number of articles by rank number, 1,…, 29, were calculated across all the 801 tables. Figure 5 shows the resulting average distribution of articles in $P_{tc}$ (to the left) into the 29 $ACPLC_t$ classes (to the right). Ranks and average number of articles across the $P_{tc}$ classes are shown for $ACPLC_t$.

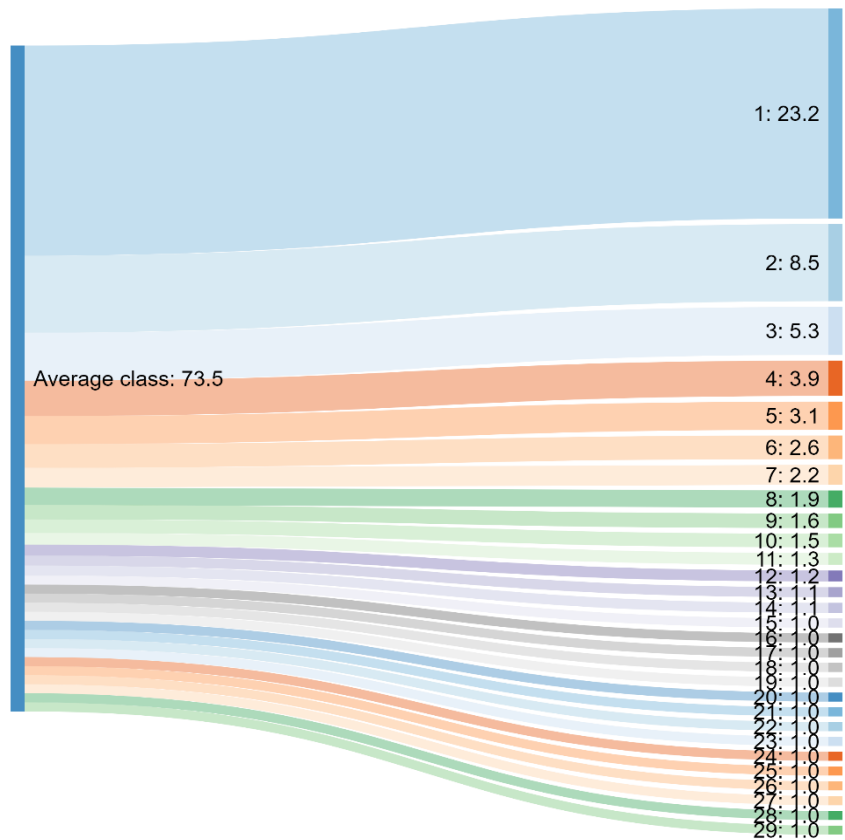

*Figure 5: Alluvial diagram for an average class. The diagram shows the distribution of reference articles in BCP$_t$ into ACPLC$_t$.*[13]

Given that we consider the classes in ACPLC$_t$ as topics, the distribution of reference articles in a typical BCP$_t$ class follows a skewed distribution of topics. About 43% of the reference articles in an average BCP$_t$ class are distributed into the two most frequent topics, and 20 topics (classes 10 to 29) are represented by a single reference article (after rounding to nearest integer). Hence, a high share of the reference articles of the average BCP$_t$ class is concentrated to a few of the ACPLC$_t$ classes. We therefore consider the match between ACPLC$_t$ and BCP$_t$ as good.

How many topics are there in ACPLC$_t$, and how large, in terms of number of Web of Science articles, is a topic? ACPLC$_t$ consists of 230,559 classes, with class sizes ranging from 2,089 to 1 publications. Figure 6 shows a histogram of the distribution of classes by class size (in terms of number of articles). Most of the classes are small in size. 93,620 classes contain less than 50 articles, and hence, 136,939 classes contain 50 articles or more. However, small classes contain a low proportion of the total number of articles in *P*. Classes with less than 50 articles constitute only approximately 4.3% of the total number of articles in *P* and classes with less than 30 articles constitute only 1.4 % of the total number of articles in *P*. The properties of the upper part of the distribution (the number and size of large classes) are not visible in the histogram of Figure 6. However, in Figure 7, size of classes have been plotted by rank order for ACPLC$_t$, (= ACPLC_3), as well as for as ACPLC_2 and as ACPLC_4. A log-10 scale is used on both the vertical axis (showing class size by number of articles) and the horizontal axis (showing ranks). For instance, the figure shows that for ACPLC$_t$, about 500 classes contain at least 1,000 articles and that about 10,000 classes contain at least 500 articles. The size of classes is dropping rather slowly, regardless of classification. The increasing granularity–from ACPLC_2 via ACPLC$_t$ to ACPLC_4–is reflected by, for example, corresponding, decreasing intercepts.

---

[13] http://sankeymatic.com/ has been used for the illustration.

Neither Figure 6 nor Figure 7 reflects the class size most articles in *P* are associated with. Because of this, we generated a weighted distribution for $ACPLC_t$ to express properties of this kind. Each class was assigned a weight equal to the number of articles it contains. Figure 8 shows a histogram of the weighted distribution. The area of this histogram reflects the number of articles assigned to classes of different size, e.g. all classes with about 600 articles per class contain about 200,000 articles. The weighted distribution have also been used to calculate the mean and median of the distribution, as well as the 10$^{th}$ and 90$^{th}$ percentiles. Resolution parameters and ARI values for each of the ACPLC_*iP*'s are reported in

| Denotation | Resolution | ARI value | # classes with # articles ≥ 50 | Weighted class size distribution measures regarding ACPLC_*i* (*i* = 1, …, 6): Mean, Median, 10$^{th}$ and 90$^{th}$ percentile (denoted $P_{10}$ and $P_{90}$) | | | |
|---|---|---|---|---|---|---|---|
| | | | | Mean # articles per class | Median # articles per class | $P_{10}$ | $P_{90}$ |
| ACPLC_1*P'* | 0.00005 | 0.132 | 59,370 | 993 | 873 | 223 | 1.913 |
| ACPLC_2*P'* | 0.00010 | 0.147 | 104,640 | 522 | 450 | 110 | 1.029 |
| ACPLC_3*P'* | 0.00015 | 0.148 | 136,939 | 357 | 305 | 75 | 716 |
| ACPLC_4*P'* | 0.00020 | 0.145 | 159,245 | 273 | 230 | 58 | 551 |
| ACPLC_5*P'* | 0.00025 | 0.139 | 174,323 | 221 | 184 | 48 | 448 |
| ACPLC_6*P'* | 0.00030 | 0.134 | 184,923 | 186 | 153 | 41 | 379 |

Table 2, together with class size distribution measures of the corresponding ACPLC_*iP*'s. The same measures are expressed in Table 3 for $ACPLC_t$ and per year, for the most recent complete ten year period, 2006-2015. In the remainder of this section, the mean, median, 10$^{th}$ or 90$^{th}$ percentile refer to weighted distributions.

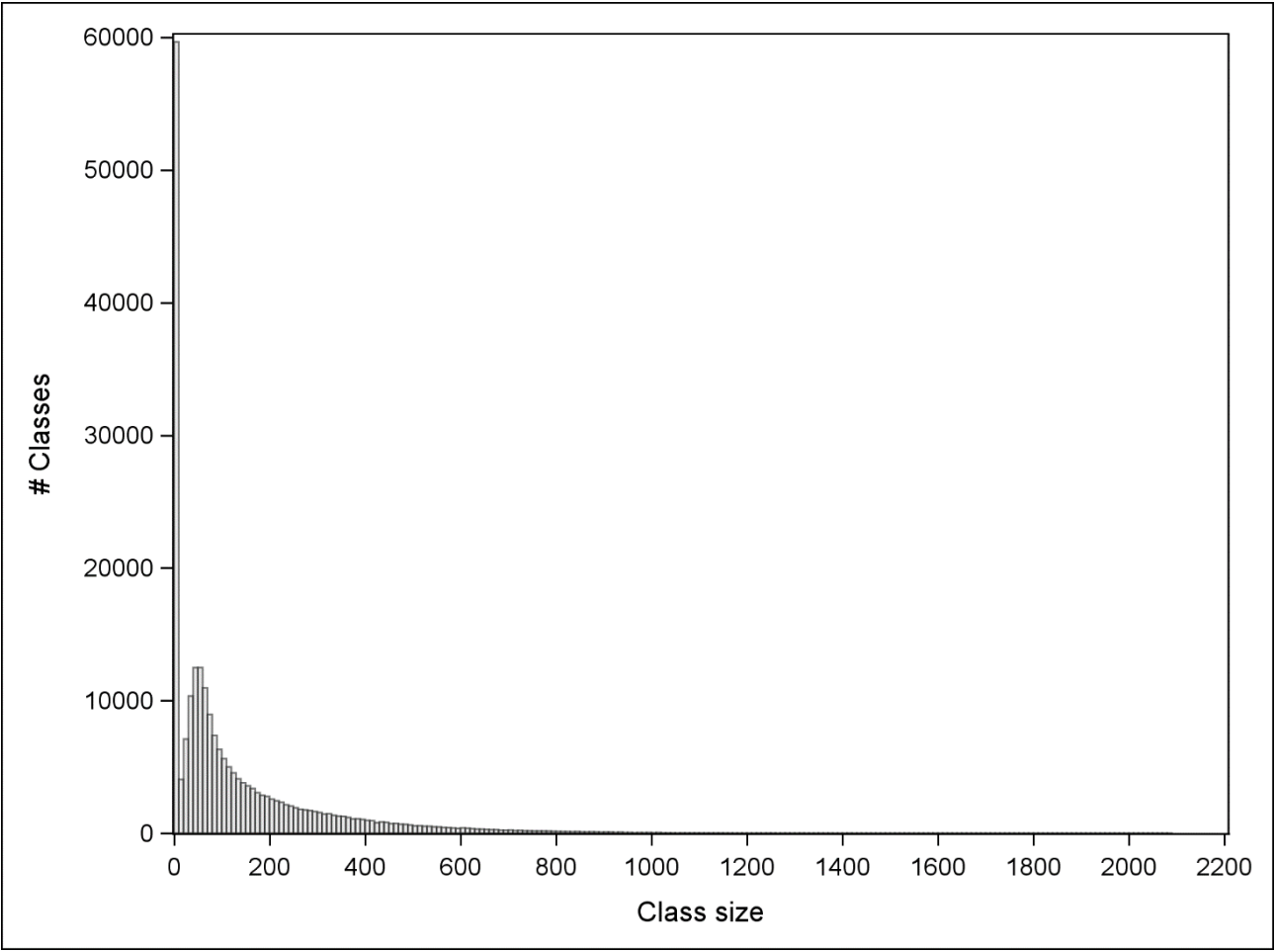

*Figure 6: Histogram of number of classes by class size for $ACPLC_t$.*

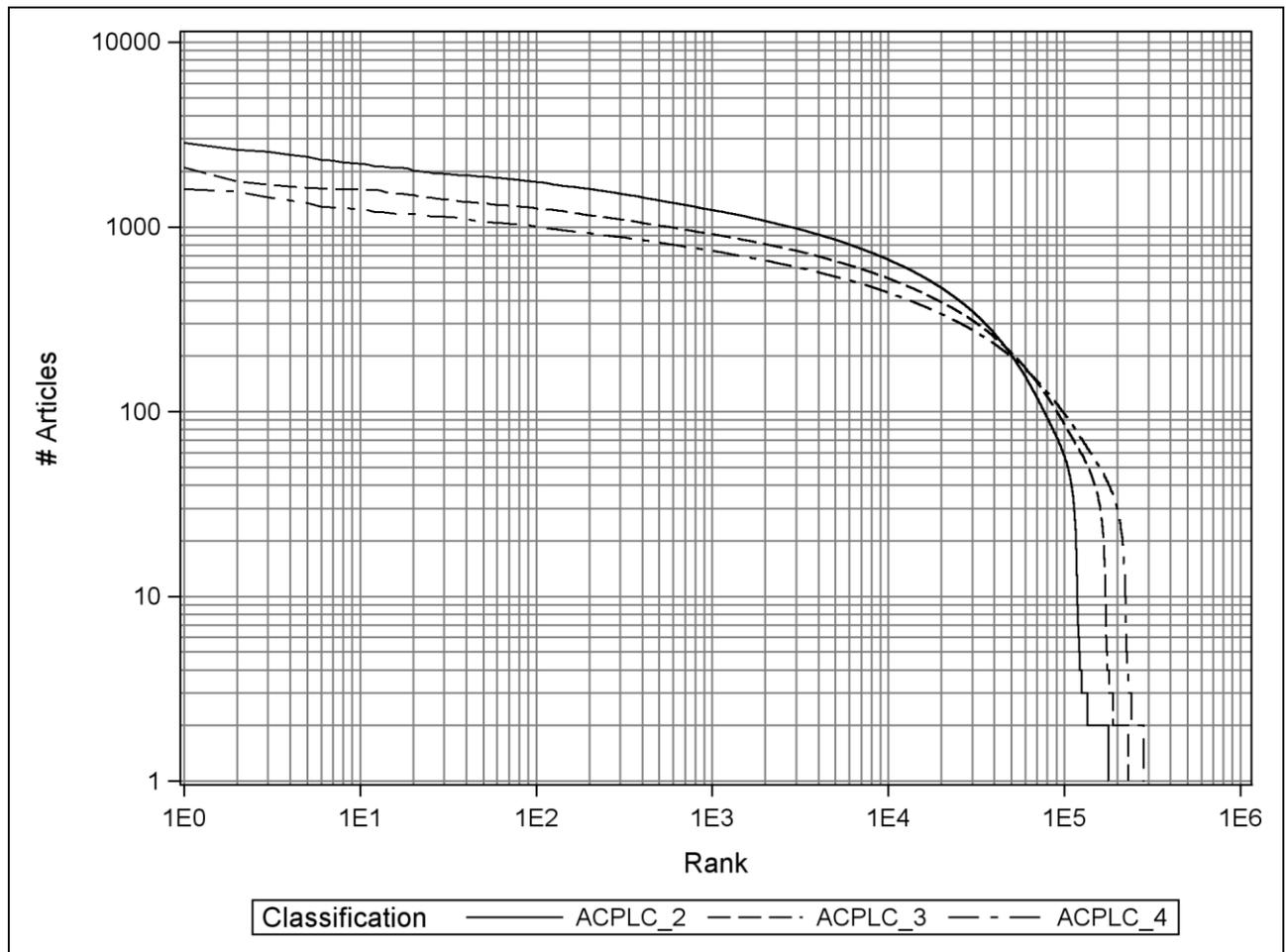

*Figure 7: Distribution of number of articles by class size for three classifications. The classes in ACPLC_2, ACPLC_3 = ACPLC$_t$ and ACPLC_4 are ordered descending by size with respect to the horizontal axis. Log-10 scale used for both axes.*

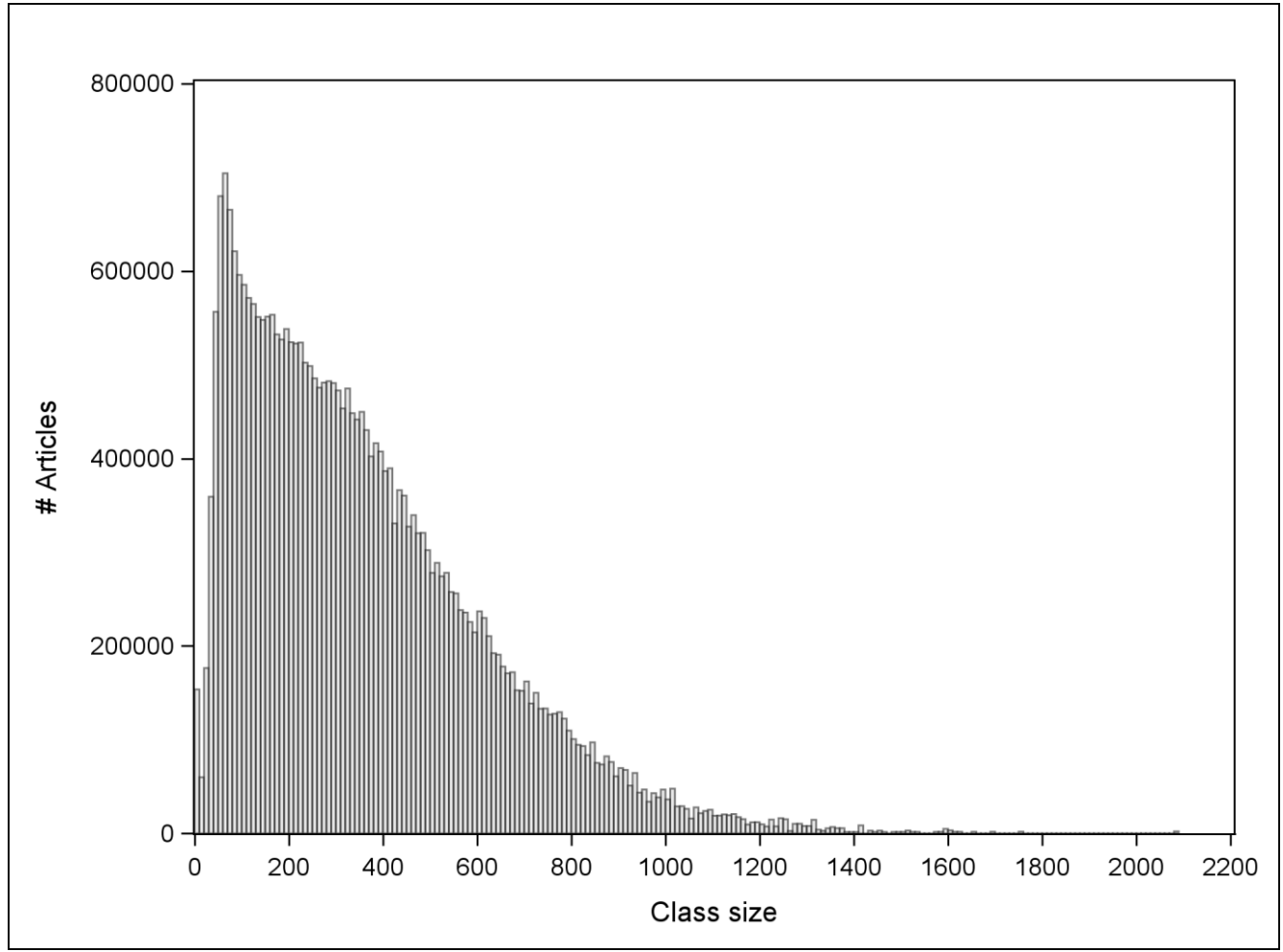

Figure 8: Histogram of number of articles by class size for $ACPLC_t$.

| Denotation | Resolution | ARI value | # classes with # articles ≥ 50 | Mean # articles per class | Median # articles per class | $P_{10}$ | $P_{90}$ |
|---|---|---|---|---|---|---|---|
| | | | | Weighted class size distribution measures regarding $ACPLC\_i$ ($i$ = 1, …, 6): Mean, Median, 10th and 90th percentile (denoted $P_{10}$ and $P_{90}$) | | | |
| ACPLC_1$P'$ | 0.00005 | 0.132 | 59,370 | 993 | 873 | 223 | 1.913 |
| ACPLC_2$P'$ | 0.00010 | 0.147 | 104,640 | 522 | 450 | 110 | 1.029 |
| ACPLC_3$P'$ | 0.00015 | 0.148 | 136,939 | 357 | 305 | 75 | 716 |
| ACPLC_4$P'$ | 0.00020 | 0.145 | 159,245 | 273 | 230 | 58 | 551 |
| ACPLC_5$P'$ | 0.00025 | 0.139 | 174,323 | 221 | 184 | 48 | 448 |
| ACPLC_6$P'$ | 0.00030 | 0.134 | 184,923 | 186 | 153 | 41 | 379 |

Table 2: For each ACPLC_i$P'$, the ARI value between ACPLC_i$P'$ and $BCP_t$, and the value of the resolution parameter used to obtain ACPLC_i, are shown, as well as number of classes with at least 50 articles and class size distribution measures for ACPLC_i.

| Publication year | # Articles | Weighted distribution measures regarding ACPLC$_t$: Mean, Median, 10$^{th}$ and 90$^{th}$ percentile (denoted $P_{10}$ and $P_{90}$) | | | |
| --- | --- | --- | --- | --- | --- |
| | | Mean # articles per class | Median # articles per class | $P_{10}$ | $P_{90}$ |
| 2006 | 989,420 | 17 | 13 | 3 | 36 |
| 2007 | 1,040,026 | 18 | 14 | 3 | 38 |
| 2008 | 1,115,118 | 19 | 15 | 3 | 41 |
| 2009 | 1,166,665 | 20 | 16 | 4 | 43 |
| 2010 | 1,210,495 | 22 | 16 | 4 | 46 |
| 2011 | 1,290,309 | 24 | 18 | 4 | 51 |
| 2012 | 1,358,175 | 26 | 19 | 4 | 56 |
| 2013 | 1,435,835 | 29 | 21 | 4 | 63 |
| 2014 | 1,478,273 | 31 | 22 | 5 | 69 |
| 2015 | 1,524,010 | 35 | 23 | 5 | 76 |

*Table 3: For the most recent complete 10 year period, the table shows class size distribution measures for ACPLC$_t$.*

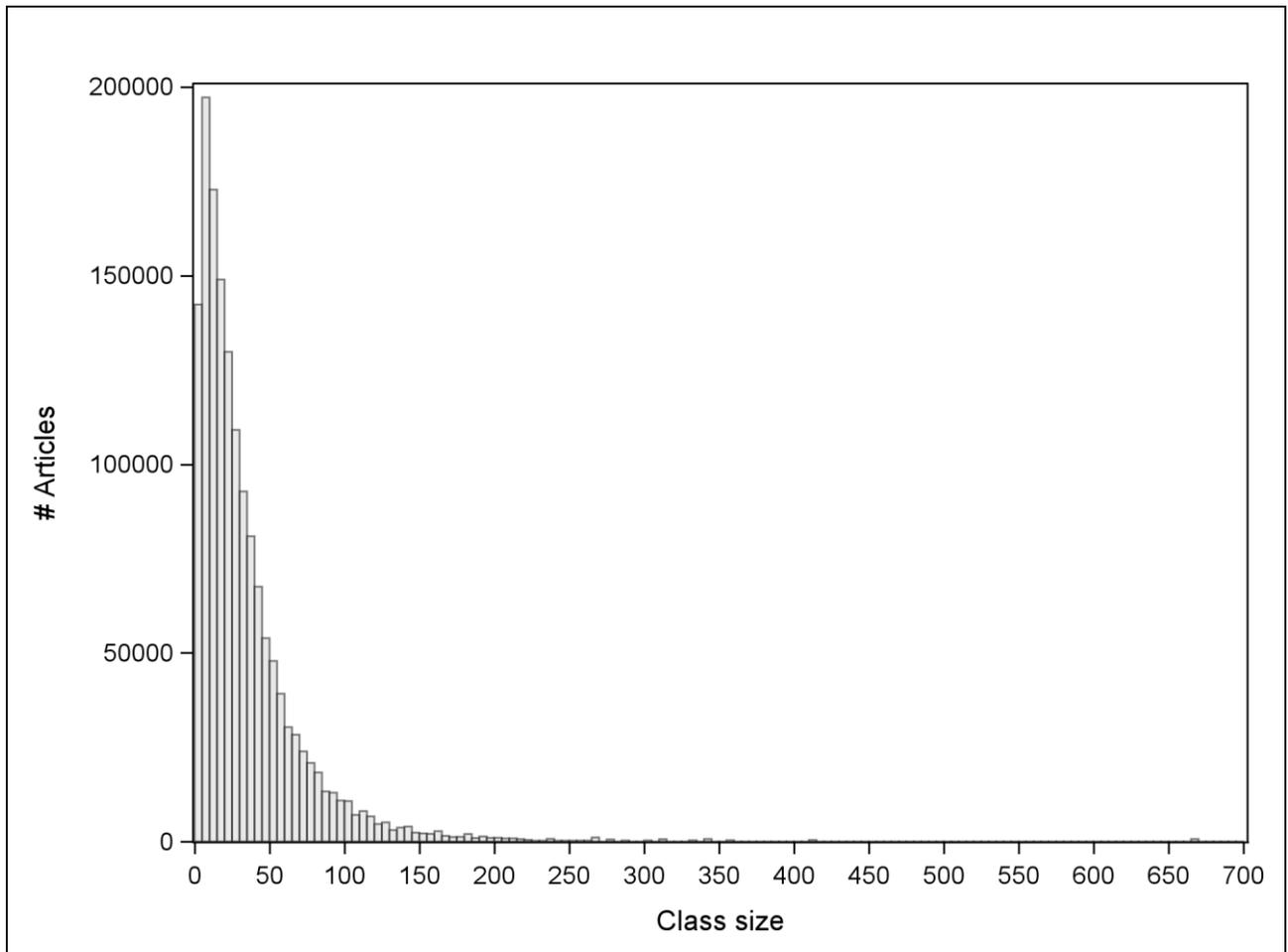

*Figure 9: Histogram of number of articles by class size, for the publication year 2015 and for ACPLC$_t$.*

For a randomly selected article *a*, it is most probable that the size of the topic class in ACPLC$_t$ to which *a* belongs is 60-70 articles (cf. the highest bar of the histogram in Figure 8). However, since the distribution is positively skewed, it is much more likely that *a* addresses a topic that is larger than 60-70 articles than smaller than this class size. In fact, the 10[th] percentile is 75 (Table 2), indicating that 90% of the articles address topics with at least 75 articles, and the 90[th] percentile is 716, hence 80% of the articles address topics consisting of 75-716 articles. The median value of $ACPLC_t$ is 305.

It can be questioned if classes with a very low number of publications, less than e.g. 50, should be considered as topics. Theoretically, a topic can be addressed by a single publication or a small number of publications. However, small classes can also be the artifact of few citation relations and the clustering algorithm used to obtain the classification. There are practical reasons to reassign small classes so that they are merged with classes with a minimum number of publications, where the minimum value is set at each granularity level of the classification (Waltman & van Eck, 2012). Future research may investigate if small classes can be considered as topics, or if they have been obtained as an artifact of the methodology.

The number of articles contributing to a topic in 2015 (the most recent complete year) is between 5 and 76, given that we only take the mid 80% of the distribution into account (Table 3 and Figure 9). The median class size is 23. The mean number of articles per topic class is growing approximately linearly across the 10 years (Table 3). This can be expected, considering the linear growth of research publication output in Web of Science.

## 4.2. The case of Journal of Informetrics

The first case used to explore if the topics obtained by the described methodology make intuitive sense concerns the topics addressed by articles in Journal of Informetrics. We choose JOI since JOI publishes articles within our field of expertise, making it possible for us to review the results and evaluate the meaningfulness of the obtained article classes and their correspondence with topics. All JOI articles in $P$ were extracted, a total of 632 articles. Let this set of articles be $P_{JOI}$. The distribution of the articles in $P_{JOI}$ into classes in $ACPLC_t$ were calculated. The 10 most frequent classes, with respect to number of $P_{JOI}$ articles, are shown in Table 4, sorted descending according to their frequency. For each class, labels were created based on author keywords in all of the articles in $P$ belonging to the class. Chi-square was used to quantify the relevance of author keywords in each class (Manning, Raghavan, & Schütze, 2008). In this implementation, the chi-square test takes into account (1) the frequency of publications in a class that contains an author keyword and (2) the expected frequency of publications in a class that contains the author keyword. The three most relevant author keywords were used to create a label for the class. We then browsed through the titles of the articles of the top 10 JOI classes (including non-JOI articles) to distinguish the topics addressed by the articles in the classes. Based on this procedure and the labels created for the classes, short topic descriptions of the 10 classes were manually added. In Table 4, labels, together with abbreviations of them, ranks, short descriptions and numerical data are shown for the classes.

Several topics can be clearly distinguished in the list of top 10 classes, e.g. regarding the three classes at the top of the list (ranks within parentheses): (1) researcher level citation indexes, (2) normalized citation indexes and (3) research mapping and classification. Other topics that can be easily identified from the class labels are (8) citation databases and (9) altmetrics.

The labeling approach we used worked well for most of the classes. However, two classes (6 and 10) did not get a label that made it easy to interpret the topic of the class. Class (6) "AUTHOR RANKING//HIGH QUALITY MANUSCRIPTS//RANKING OF AUTHORS" mainly contains articles that use network-based methods for ranking of authors (and sometimes other entities). PageRank is mentioned in many article titles. Other terms that occur in the titles, and bearing witness of the topic orientation of this class, is "network flows", "graph-based algorithms", "network structure", "network model" and "centrality measure". The articles in class (10) "UNCITEDNESS FACTOR//WORLD JOURNAL OF GASTROENTEROLOGY//CITATION HISTORIES" address the life time of articles and dynamics of citation uptake. Terms found in the titles include "citation growth", "citation age", "uncitedness" and "sleeping beauties".

Three classes, (4) "RESEARCH COLLABORATION//SCIENTIFIC COLLABORATION//CO AUTHORSHIP" ("intCollab"), (5) "AUTHOR CO CITATION ANALYSIS//BIBLIOGRAPHIC COUPLING//CO CITATION ANALYSIS" ("citMap") and (7) "CO AUTHORSHIP NETWORKS//SCIENTIFIC COLLABORATION//CO AUTHOR NETWORKS" ("authNet") are similar in scope. However, when browsing the titles of the articles we found that the classes can be differentiated based on the topical orientation of the articles in the classes (or at least a core set of articles in each class). Nevertheless, some of the articles within these classes would fit within two or more of the classes.

The articles in the citMap class focus on citation measures for mapping and visualization. E.g. the term "co-citation" is much more frequent in the titles in this class (16% of articles include this term) than in the authNet and intCollab, classes in which the term is absent. The articles within the authNet class focus on research collaboration and researcher networks. The term "network" occurs in 66% of the article titles in this class, compared to 11% in the citMap class and 9% in the intCollab class. The intCollab class is also focused on research collaboration. Compared to the authNet class, the focus of the articles in the intCollab class is at a higher level of aggregation, countries rather than individual researchers, and many of the articles apply comparative approaches. The focus of this class is also more oriented towards international collaboration: the term "international" is much more frequent in the titles in this class (25%) than in the citMap (1%) and authNet (3%) classes.

In conclusion, this case shows that topics within JOI, that make intuitive sense to us as field experts, can be identified by use of the classes in $ACPLC_t$. Further, each of the classes we studied addresses a distinct topic and the overlap of topics between classes is rather low.

| Rank | Class-id | # Articles | Share articles in JOI (%) | Abbreviation | Label based on author keywords (top 3 ranked by $chi^2$) [Manually added short topic description in brackets] |
|---|---|---|---|---|---|
| 1 | 6741 | 117 | 18.5 | Hind | H INDEX//HIRSCH INDEX//G INDEX [Researcher level citation indexes] |
| 2 | 11564 | 112 | 17.7 | normInd | FIELD NORMALIZATION//SOURCE NORMALIZATION//RESEARCH EVALUATION [Normalized citation indexes] |
| 3 | 24854 | 26 | 4.1 | mapMeth | OVERLAY MAP//SCIENCE OVERLAY MAPS//JOURNAL CLASSIFICATION [Research mapping and classification] |
| 4 | 9340 | 24 | 3.8 | intCollab | RESEARCH COLLABORATION//SCIENTIFIC COLLABORATION//CO AUTHORSHIP [Research collaboration with focus on international collaboration] |
| 5 | 14932 | 23 | 3.6 | citMap | AUTHOR CO CITATION ANALYSIS//BIBLIOGRAPHIC COUPLING//CO CITATION ANALYSIS [Citation measures for mapping of bibliometric networks] |
| 6 | 50743 | 20 | 3.2 | articleLife | UNCITEDNESS FACTOR//WORLD JOURNAL OF GASTROENTEROLOGY//CITATION HISTORIES [Lifetime of articles, e.g. sleeping beauties] |
| 7 | 39166 | 17 | 2.7 | authNet | CO AUTHORSHIP NETWORKS//SCIENTIFIC COLLABORATION//CO AUTHOR NETWORKS [Research collaboration with focus on networks] |
| 8 | 36941 | 15 | 2.4 | citDab | GOOGLE SCHOLAR//SCOPUS//WEB OF SCIENCE [Citation databases] |
| 9 | 51930 | 14 | 2.2 | altMet | ALTMETRICS//MENDELEY//RESEARCHGATE [Altmetrics] |
| 10 | 76509 | 13 | 2.1 | authRank | AUTHOR RANKING//HIGH QUALITY MANUSCRIPTS//RANKING OF AUTHORS [Author rankings] |

*Table 4: Distribution of reference articles published in JOI into classes in $ACPLC_t$. Ten most frequent classes, with respect to number of $P_{JOI}$ articles, are included.*

## 4.3. The case of nanocelluloses

The second case used to explore if the topics obtained by the described methodology make intuitive sense concerns nanocelluloses. There are several reasons to choose nanocelluloses as a subject area for this test. First, we want to test the results outside the subject area in which we are active. Second, the nano area in general, and the area of nanocelluloses in particular, is of interest, since it is an area of recent emergence and is, in terms of publication output, growing rapidly. The nano area has been studied in numerous scientometric studies. A class containing articles reporting scientometric studies on the nano area could be identified in ACPLC$_t$ (e.g. Hullmann & Meyer, 2003; Porter, Youtie, Shapira, & Schoeneck, 2008; Schummer, 2004). This class contains 285 articles, including three articles mentioning nanocelluloses in their titles (Milanez, do Amaral, Lopes de Faria, & Rodrigues Gregolin, 2013, 2014; Milanez et al., 2016). A third reason to study nanocelluloses is that we are somewhat familiar with this subject area, since we have studied it in our practice on demand of a client.

We used a review article as the point of departure for this case. The review is titled "Nanocelluloses: A New Family of Nature-Based Materials" and treats the preparation and use of three types of nanocellulose: (1) microfibrilliated cellulose, (2) nanochrystalline cellulose and (3) bacterial cellulose (Klemm et al., 2011). The review is identified as a highly cited paper in Web of Science and has, at the time of writing, been cited more than a thousand times. It is the most cited paper retrieved by a Web of Science topic search on "nanocellulose*", where the truncation operator "*" stands for a group of zero or more character occurrences.

The review contains 391 cited references. 227 of these have been assigned to a class in ACPLC$_t$. Let this set of articles be $P_{nc}$. The discrepancy between the number of references and the number of articles in $P_{nc}$ is mainly accounted for by references to publication types other than articles (foremost conference papers and patents), references to articles published in sources not covered by Web of Science and incomplete or erroneous references.

Table 5 shows how the articles in $P_{nc}$ are distributed into the six most frequent classes in ACPLC$_t$ (classes with more than two articles in $P_{nc}$). The table is sorted descending by frequency. Labels for the classes were constructed with the same methodology as in the JOI case. The top three classes are clearly dominant and cover 75% of the articles in $P_{nc}$ (Figure 10). The labels of these classes correspond to the three types of nanocelluloses outlined by the review article. The topics of the fourth and sixth ranked classes are relatively easy to distinguish. The rank 4 class, labeled "CELLULOSE I BETA//CELLULOSE I ALPHA//CELLULOSE", treats the chemical structure of cellulose, and the rank 6 class, labeled "CELLULOSE MODEL SURFACE//CELLULOSE THIN FILMS//CELLULOSE MODEL FILMS", treats the elaboration of cellulose thin films. The rank 5 class has a similar label as the rank 1 class and, by looking at labels only, these classes appear to overlap regarding their topic content. By examining titles, journals and journal categories, we can, however, distinguish between the topics addressed by the articles within these two classes. The rank 1 class is oriented towards basic research about cellulose nanocrystals, while the rank 5 class is oriented towards applications of such crystals. This is manifested by a higher share of the articles in the rank 1 class being located in the Web of Science subject categories "Chemistry, Multidisciplinary" and "Materials Science, Multidisciplinary", compared to the rank 5 class. The rank 5 class contains a higher share of articles within the Web of Science subject categories "Chemistry, Applied", "Agricultural Engineering" and "Agronomy" than the rank 1 class. The distinction between the two classes is also manifested by the occurrences of terms associated with the materials used for the elaboration of nanocelluloses. Examples of such terms are "cotton", "sugarcane bargasse", "rye straw", "banana fibers" and "agricultural waste". Such terms occur much more frequently in the titles of the articles in the rank 5 class.

The three types of nanocelluloses have also been identified in previous bibliometric studies of nanocelluloses (Milanez et al., 2014, 2016). The results of the study of Milanez et al. (2016) is interesting in comparison with our results. These authors used an ACPLC developed and published by Waltman and van Eck (2012) to identify topics, and identified classes within the subject area of nanocelluloses in two steps. First they retrieved publications from Web of Science by a topic search on terms associated with nanocelluloses. In a second step, they calculated the frequency distribution of the retrieved set of publications into classes in the ACPLC. They identified two main classes in which 44.9% of the retrieved publications were included. The first class includes both microfibrilliated cellulose (nanofibrils) and nanochrystalline cellulose (cellulose nanocrystals). The second class treats bacterial nanocellulose. Thus, two of the three types of nanocelluloses are associated with the same publication class. This indicates that the granularity of the ACPLC Milanez et al. used is too coarse to accurately correspond to topics.

In conclusion, this case shows that our approach has resulted in an ACPLC (ACPLC$_t$) that contains classes for each of the three types of nanocelluloses. Bacterial cellulose and microfibrilliated cellulose are located in one class each. Basic and applied nanochrystalline cellulose research are separated into two classes. The outcome indicates that ACPLC$_t$ is useful for identifying topics within nanocelluloses. Additionally, ACPLC$_t$ can also be used to retrieve articles within each of the identified classes.

| Rank | Author keywords (3 top ranked by chi$^2$) | # reference articles in class | Share of total # reference articles that have been assigned to the class (%) |
|---|---|---|---|
| 1 | CELLULOSE NANOCRYSTALS//CELLULOSE WHISKERS//CELLULOSE NANOCRYSTALS CNCS | 81 | 36 |
| 2 | MICROFIBRILLATED CELLULOSE//NANOFIBRILLATED CELLULOSE//CELLULOSE NANOFIBRILS | 46 | 20 |
| 3 | BACTERIAL CELLULOSE//ACETOBACTER XYLINUM//GLUCONACETOBACTER XYLINUS | 44 | 19 |
| 4 | CELLULOSE MODEL SURFACE//CELLULOSE THIN FILMS//CELLULOSE MODEL FILMS | 4 | 2 |
| 5 | CELLULOSE NANOCRYSTALS//NANOCELLULOSE//CELLULOSE NANOFIBERS | 3 | 1 |
| 6 | CELLULOSE I BETA//CELLULOSE I ALPHA//CELLULOSE | 3 | 1 |

*Table 5: Distribution of reference articles from the review "Nanocelluloses: A New Family of Nature-Based Materials" into classes in ACPLC$_t$. Classes containing three or more reference articles have been included.*

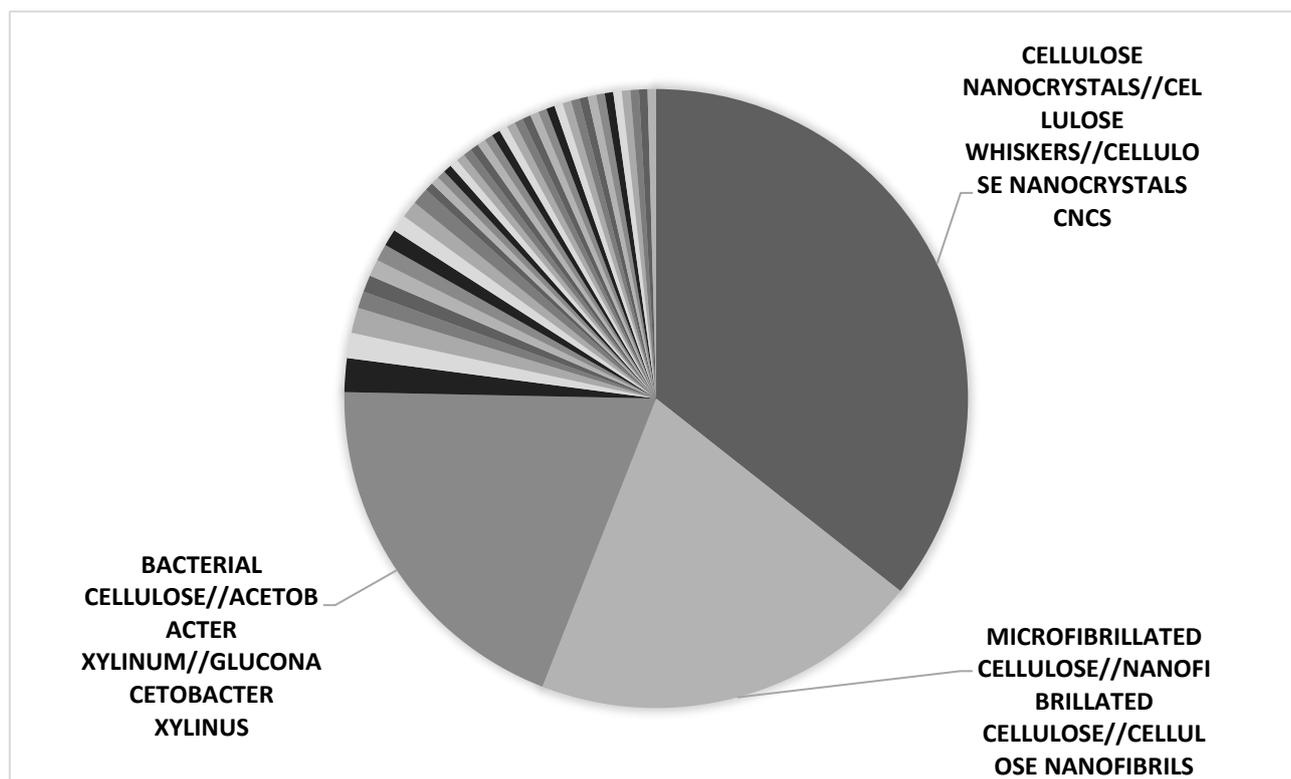

*Figure 10: Distribution of reference articles from the review "Nanocelluloses: A New Family of Nature-Based Materials" into classes in ACPLC$_t$. Labels are shown for the three largest classes.*

# 5 Conclusions

In this study we have discussed how the resolution parameter given to the Modularity Optimizer software can be calibrated so that obtained publication classes correspond to the size of topics. Synthesis publications have been used as a baseline for the calibration. The underlying assumption of our approach is that synthesis publications in general address a topic. By measuring the similarity between (1) the baseline classification and (2) multiple classifications obtained by using different values of the resolution parameter, we have identified a classification whose granularity corresponds to topics.

Šubelj et al. (2016) point out that the difference in size of the classes of an ACPLC should not be too large and that, for practical reasons, "the number of very small clusters should be minimized as much as possible". It can be expected that some topics are larger in size than others. In the ACPLC which best correspond to topics in this study (ACPLC$_t$), 80% of the articles address topics consisting of 75-716 articles and classes with 30 articles or less constitute only approximately 1.4% of the total number articles in *P*. The distribution follows a typical scientometric distribution, and we therefore consider the results, regarding class sizes, as satisfying. Further, the two case studies indicate that the classes make intuitive sense, that a topic for each class can be identified and that the topical difference between classes can be distinguished. Still, there is some overlap between topics. Considering the topics of the articles that we have studied more closely in the case studies, some articles address topics of more than one class or a topic in the borderland between two or more classes. It is a disadvantage of the used approach that publications cannot be assigned to more than one class. Classifications that allow publications to be assigned to several classes might be an alternative. However, such approaches also give rise to some issues: (1) Of practical reasons, there is a need to limit the maximum size of classes, an issue that is similar to the issue addressed in the present paper. (2) It is likely that the number of classes to which a publication can be assigned needs to be limited. (3) The approach causes multiple counting of some publications (but not of others) when full count statistics are compiled. In view of these issues, we consider a logical classification to be of more practical use, at least for some analytical purposes.

An issue that needs further attention is how well algorithmic classifications manage to deal with differences regarding size of topics in different subject areas. Does the methodology used in this study result in classes that are perceived as topics in all subject areas, or are the classes perceived as topics in some subject areas, while perceived as more broad or more narrow than topics in other subject areas? This is a question to be answered by future research.

We have looked into the granularity of an ACPLC at one level of granularity only. Several levels are needed to create an ACPLC that is to be used for a wide range of bibliometric analyses. We are planning to address the level of specialties in a future study. Other levels to study in future work might be the level of research disciplines, which we consider as broader than specialties, and the level of broad subject categories, which we consider to be the most coarsely grained level of an ACPLC. A related issue, studied in the literature, is the granularity of classifications used for field normalization of citation rates (Perianes-Rodriguez & Ruiz-Castillo, 2017). The levels obtained by our approach, or by future similar approaches, may not be optimal for normalization. Normalization could therefore be done with other techniques, e.g. Colliander (2015) or Waltman and van Eck (2013b). How the granularity levels of a standard ACPLC relate to citation normalization is something we consider to be of interest for future research.

A weakness of our approach is that it is not easily repeated. The resolution parameter needs to be set differently if the underlying data is not exactly the same. When data is updated, new publications and their references will influence the network, and the resolution parameter may need to be adjusted. However, the results of this study may guide adjustments of the resolution parameter, since approximate sizes of topics have been outlined.

## Acknowledgments

We would like to thank three anonymous reviewers for their comments on an earlier version of this paper.

## Appendix: Definition of the Adjusted Rand Index

Let *X* and *Y* be two partitions of a set *W* with *n* objects. *Adjusted Rand Index with respect to X and Y*, ARI(*X*, *Y*), is then defined as follows (Hubert & Arabie, 1985):

$$\mathrm{ARI}(X,Y) = \frac{\sum_{ij}\binom{n_{ij}}{2} - \left[\sum_i \binom{a_i}{2} \sum_j \binom{b_j}{2}\right] / \binom{n}{2}}{\frac{1}{2}\left[\sum_i \binom{a_i}{2} + \sum_j \binom{b_j}{2}\right] - \left[\sum_i \binom{a_i}{2} \sum_j \binom{b_j}{2}\right] / \binom{n}{2}} \tag{4}$$

where $n_{ij}$ is the number of objects that belong to the *i*th class in *X* and the *j*th class in *Y*, $a_i$ the number of objects in the *i*th class of *X*, and $b_j$ the number of objects in the *j*th class of *Y*.

In our case, *W* corresponds to *P'*: the union of the classes of the baseline classification BCP$_t$. Thus, the *n* considered objects are the 2,786,203 reference articles underlying BCP$_t$. Further, regarding correspondents to *X* and *Y*, for each of the six calculations of ARI, one of the two involved partitions is BCP$_t$ and the other is ACPLC_*i*P' (*i* = 1, ..., 6), derived from ACPLC_*i*, the *i*th algorithmically constructed publication-level classification.